# Speculation On A Source of Dark Matter

Richard Packard
Physics Department, university of California, Berkeley CA 94720

By drawing an analogy with superfluid $^4$He vortices we suggest that dark matter may consist of irreducibly small remnants of cosmic strings.

As experiments have increasingly eliminated known particles as dark matter candidates one is tempted to imagine more exotic entities to explain the astronomical observations that suggest the presence of missing mass. This note presents, in a simple outline, a suggestion that dark matter may consist of very small entities that are the final irreducible remnants of cosmic strings. We call them talons. Talons are in a metastable state and cannot interact inelastically to give energy to any entity. They may thus be undetectable in any conventional search.

Although this note is very speculative the reader may need reminding that many ideas related to cosmic strings are also speculative. There are already hundreds of theoretical papers concerning cosmic strings[1] in spite of there being, after 40 years, no observational evidence for their existence. Additionally there are fairly strong observational constraints[2] suggesting that filamentary cosmic strings did not persist much after the initial instants of the big bang [3]. Since this note may disagree with the conclusions developed by others, the author pleads indulgence, as his field of study is superfluid vortex lines, not cosmic strings. However if the ideas expressed here show some connection to nature, others more knowledgeable may be stimulated to develop them further.

We make analogy with the known properties of quantized vortices in superfluid helium[4,5]. Various processes create these vortices. They are like simple classical vortices except that the fluid circulation around a normal core is quantized in units of h/m where h is Plank's constant and m is the helium atomic mass. The irreducible core size, which has only radial structure, is of atomic dimensions. Due to energy considerations, the lines are normally single quantized

The equations of motion of quantized vortex filaments are known and give rise to complex motion predictable via simulations[6]. For the case of a pure superfluid, given some initial simple configuration of lines, future states become complex tangles of spaghetti-like vorticity. When two line segments get very close there are reconnection[7] events. Then the lines break into shorter segments with some energy carried way via acoustic radiation[8]. When a line reconnects on itself, a vortex ring or loop will form[9]. The ring travels away carrying some of the energy of the original line. Those rings that do not intersect with a line may eventually hit a bounding wall and vanish. If no energy is added to the system the energy of the initial lines will continually dissipate as the moving vortex rings and loops encounter walls.

The vortex tangle will cascade into rings whose size eventually reaches the core dimensions. At this point they can shrink no more and may be considered as elementary excitations of the liquid. The superfluid excitations called rotons have been referred to as "the ghost of a vanishing vortex ring". Since a real superfluid is bounded by the walls of an enclosure, the elemental rings eventually are absorbed as heat via collisions with the walls[10]. The cascade to smaller and smaller loops seems inevitable although computer simulations cannot follow the process to core-size length scales.

The key point here is that any collection of simple vortices must eventually cascade into core size rings. This may be a self-organizing property of nonlocally interacting filamentary structures that permit reconnections.

We now join this discussion to the properties of cosmic strings[11], entities first predicted by Thomas Kibble[12]. In the first moments after the big bang, the rapidly cooling universe undergoes a series of symmetry breaking transitions, conceptually similar to transition-related features in several condensed matter systems. In such rapid phase transitions, a common feature is to freeze in some topological line (i.e. string-like) defects. These are referred to as cosmic strings. The simplest cosmic strings are analogous to quantized vortices in superfluid helium. However there are wide variations in the conjectured structure and physical properties of the strings, mainly depending on the energy scale wherein they form. For example when grand unified theory (GUT) interactions segue to an epoch of lower symmetry, the strings are thought to be characterized by a core dimension as small as $10^{-27}$cm and a lineal mass density as large as $10^{22}$gm/cm.

We wish to restrict this discussion to the simplest possible cosmic strings, those most analogous to the helium quantized vortices. In particular, we assume the strings have no internal structure other than some radial function whose properties can be characterized by the core size. The core size would be irreducible in our model.

The equations of motion of cosmic strings are believed to be known. Simulations[13] that start with a few strings show that they would quickly evolve into a complex tangle of cosmic spaghetti. The simulations of the string distributions look strikingly similar to the quantized vortex simulations. It is this simple observation that provoked the model we propose here.

Reconnection events occur when two string filaments cross with some of the energy possibly radiated away as gravity waves. When cosmic strings cross on themselves loops are formed. Analogous to the superfluid vortices, it seems likely that the initial strings will eventually cascade into ever smaller loops. **Our speculation is that there is an end point of any energy dissipation mechanism for a string with an irreducible core.** This seems inescapable if the core is irreducible. Ultimately any internal motion will be frozen out and dissipation will cease.

Some cosmic ring structures have been examined theoretically with possible stability being a central question. Such loops or rings have been called vortons[14]. These have been based on models of superconducting strings whose form might or might not be stabilized by angular momentum. For strings with internal core structure more complex than simply radial these vortons may eventually dissipate and would not be a component of dark matter. At least one analysis[15] indicates that vortons with thick cores can be stable.

We restrict our speculation to strings with no internal structure and assume that, like $^4$He vortices, their terminal configuration is an entity whose diameter is comparable to the core size.[16] For discussion purposes we will name these smallest irreducible structures, "talons".

Since talons are already as small as possible (i.e. the size of the core), they cannot decay in a continuous manner and are therefore metastable. Unlike the superfluid case where walls provide a place for the small rings to vanish, the universe has no "walls" so the final fate of the talons is to form a distribution of elemental mass-carrying entities. The talons may interact with matter gravitationally, but are unlikely to annihilate via collisions with baryonic matter. Therefore they would be very difficult to detect in conventional dark matter searches that look for events where a dark matter particle gives up its energy to some matter in a detector.

As mentioned above, the theorized internal structure and physical parameters of cosmic strings vary over many orders of magnitude, depending on what energy scale is assumed for the relevant phase transition wherein they are created. In the case of a phase transition related to the grand unified energy scale, the lineal mass density of a cosmic string has been estimated to be as high as $\sim 10^{22}$ gm/cm. For such massive strings the associated core size is conjectured to be as small as $\sim 10^{-27}$ cm. Thus, the mass of one talon might be $\sim 10^{-5}$ gms., 22 orders of magnitude more massive than a baryon. This implies that a talon number density many orders of magnitude smaller than baryon number density could play a comparable gravitational role in forming cosmic structures. With such a low density and small size an annihilation event would seem unlikely[17]. Thus the talons left behind after the big bang might persist today. These talons may be the missing mass characterized as dark matter.

In summary, arguing by analogy with superfluid vortices, we introduce the concept of a talon, **an entity that is the final, irreducible segment of a cosmic string.** The talon will have a spatial extent the size of a cosmic string core and may have enormous mass compared to ordinary fundamental particles. Thje talon may not couple to known matter by any interaction other than gravity. This would make their detection very difficult with known techniques. We suggest that talons, dispersed throughout the universe, may provide the non-luminous mass known as dark matter. If talons do exist, the universe is a quite different environment than we usually imagine

---

[1] For a recent review of cosmic strings see Copeland, E.T. and Kibble, T.W. B., Cosmic Strings and Superstrings, Proc. Roy. Soc A, doi:10.1098/rspa.2009.0591. For an earlier review see: : Vilenkin, A. and Shellard, E.P.S. *Cosmic Strings and Other Cosmological Defects,* Cambridge monographs on Mathematical Physics, (1994) Cambridge University Press

[2] Miyamoto K, and Nakayama K., Cosmological and astrophysical constraints on superconducting cosmic strings**,** arXiv:1212.6687v1 [astro-ph.CO]

[3] Jeong, E and Smoot, G.F., Search for Cosmic Strings in cosmic microwave anisotropies, Astrophysics Journal, 624, 21-27, 2005

[4] Donnelly, R.J*., Quantized Vortices in Helium II*, (Cambridge University Press, Cambrideg 1990)

[5] Barenghi, C.F. , Donnelly, R.J., Vinen W.V. ed. *Quantized Vortex Dynamics and superfluid turbulence*, Berlin: Springer Verlag, 2001, pp3-13, Introduction to superfluid vortices and turbulence:

[6] For example see: Tsubota M, Araki T, Nemirovskii SK. Dynamics of vortex tangle without mutual friction in superfluid $^4$He. Phys Rev B. 2000;62(17):11751–11762.

[7] Koplik, J. and Levine, H., Vortex reconnection in superfluid helium, Phys. Rev. Lett. **71**, 1375 (1993)

[8] Berenghi, C.F., Turbulent Dissipation near absolute zero, European J. of Mechanics B/Fluids, **23**, (2004) 415-425, Fluids

[9] Kursa M, Bajer K, Lipniacki T. Cascade of vortex loops initiated by a single reconnection of quantum vortices. Phys Rev B. 2011;83(1):014515

[10] There are more complex vortices in superfluid $^3$He that have internal degrees of freedom. These rings can possibly dissipate without requiring a boundary.

[11] Ref. 1

[12] Kibble T.W.B. (1976) Topology of Cosmic Domains and Strings J. Phys. **A** 9, 1387-1398

[13] http://www.damtp.cam.ac.uk/research/gr/public/cs_evol.html

[14] Davis R.L., Shellard E.P.S., Cosmic Vortons, Nucl. Phys B323, 209 (1989)

---

[15] Garard J, Radu E, Volkov M.S. , Stable Cosmic Vortons, arXiv: 1303.3044v2 [hep.th] 28 Oct 2013

[16] It might be useful for the reader to consider the case of an ordinary piece of string with a loose overhand knot placed along it. If the ends of the string are pulled tight the knot decreases in size until it jams when the knot's diameter is comparable to the string's diameter.

[17] It is interesting to note that if a talon annihilates the resultant energy released is far in excess of the most energetic cosmic rays that have been detected.